\documentclass{aa}
\usepackage{graphicx}
\usepackage[varg]{txfonts}
\usepackage{natbib}
\usepackage{float}
\usepackage{color}

\begin{document}
  \title{The impact of type Ia supernovae on main sequence binary companions}

  \author{R.~Pakmor \and
          F.~K.~R\"{o}pke \and
          A.~Weiss \and
          W.~Hillebrandt
          }

  \institute{Max-Planck-Institut f\"{u}r Astrophysik, Karl-Schwarzschild-Str. 1, 85741 Garching, Germany\\
          \email{rpakmor@mpa-garching.mpg.de}
          }

  \date{Received 26 June 08; accepted 13 July 08}

  \abstract{The nature of Type Ia supernova progenitors is still
  unclear. The outstanding characteristic of the single-degenerate
  scenario is that it contains hydrogen in the binary companion of the
  exploding white dwarf star, which, if mixed into the ejecta of the
  supernova in large amounts may lead to conflicts with the
  observations thus ruling out the scenario.}{We investigate the effect of the impact of Type Ia
  supernova ejecta on a main sequence companion star of the progenitor
  system. With a series of simulations we investigate how different
  parameters of this system affect the amount of hydrogen stripped
  from the companion by the impact.}{The stellar evolution code
  GARSTEC is used to set up the structure of the companion stars
  mimicking the effect of a binary evolution phase. The impact itself
  is simulated with the smoothed particle hydrodynamics code
  GADGET2.}{We reproduce and confirm the results of
  earlier grid-based hydrodynamical simulation. Parameter studies of
  the progenitor system are extended to include the results of recent
  binary evolution studies. The more compact structure of the companion
  star found here significantly reduces the stripped hydrogen mass.}{The low
  hydrogen masses resulting from a more realistic companion structure
  are consistent with current observational constraints. Therefore,
  the single-degenerate scenario remains a valid possibility for Type
  Ia supernova progenitors. These new results are not a numerical effect,
  but the outcome of different initial conditions.}


  \authorrunning{R.~Pakmor et al.}

  \maketitle

  \section{Introduction}
    \label{sec:introduction}
    While the progenitors for Type Ib/c and Type II supernovae are
    known, Type Ia supernovae (SNe~Ia) still elude an identification of
    their progenitor system. This is an unpleasant situation given the
    fact that these objects are one of the most important
    tools to determine cosmological parameters. By virtue of empirical
    calibration methods \citep[e.g.,][]{phillips1993a} they can be
    used as  standardizable candles for distance measurements. This
    calls for an understanding of the mechanism of SNe~Ia; and indeed,
    some progress has been made in recent years in 
    understanding the explosion mechanism in terms of
    thermonuclear explosions of white dwarf (WD) stars
    \citep[e.g.,][]{reinecke2002d,gamezo2003a,roepke2005b, roepke2007b,
    mazzali2007a, roepke2007c}. In order to judge potential
    systematic errors in SN~Ia cosmology, a theoretical
    connection between the explosion characteristics and properties of
    the progenitor system would be desirable. Yet despite all
    efforts on both the theoretical and on the observational side,
    the nature of the progenitor system remains enigmatic.

    The stabilization of WDs against gravity does not depend
    on a finite energy source such as the nuclear burning in normal
    stars. Due to the Fermi pressure of a degenerate electron gas,
    single WD stars are in principle eternally stable. Thus, some
    additional dynamics is required in order to reach an explosive state.
    The most likely possibility is a WD being part of a
    binary system and accreting matter from its 
    companion.

    Current progenitor models distinguish between the \emph{single
    degenerate} and the \emph{double degenerate} scenario. The former
    \citep[proposed 
    by][]{whelan1973a} assumes a ``normal'', non-degenerate star to be
    the binary companion -- either a main sequence (MS) star or a red
    giant (RG). In this case, the WD accretes mass from its MS or RG
    companion via Roche-lobe overflow  or by winds (symbiotic
    systems) until it approaches the Chandrasekhar mass. The densities
    reached in the core of the WD are then sufficiently high
    to trigger nuclear reactions which finally cause a thermonuclear
    explosion of the star \citep[but note that an explosion before reaching
    the Chandrasekhar mass may also be possible, e.g.][]{fink2007a}. 
    The double degenerate scenario \citep{iben1984a, webbink1984a}, on the other
    hand, assumes a binary
    system of two WDs. Due to gravitational wave emission,
    the system becomes unstable at some point, the WDs merge, and may
    eventually explode
    in a SN Ia. A summary of arguments in favor of and against both
    scenarios can be found in \cite{livio2000a}. In theoretical
    modeling, the single-degenerate Chandrasekhar-mass scenario has received most attention
    recently. By constraining the amount of fuel available in the
    thermonuclear explosion, it provides a natural explanation for the
    observed uniformity of SNe~Ia. 

    Observationally, its hard to distinguish between the progenitor
    scenarios. One fundamental difference is the complete
    absence of hydrogen in the double degenerate scenario as it
    assumes a merger of two carbon/oxygen WDs. In contrast, in the
    typical single degenerate scenario, hydrogen is the
    main constituent of the companion. An exception are
    helium-accretors in which a more evolved companion
    star has lost his hydrogen envelope.
    Thus, the WD accretes helium instead of hydrogen and hydrogen is
    missing in
     the system. \citet{kato2003a} reported a possible detection of
    such an object.
    In the standard scenario, however, the companion star features a
    hydrogen 
    envelope; and at least some part of it is expected to be
    carried away by the SN~Ia ejecta impacting the
    companion. This, in principle,
    causes a problem for the single-degenerate scenario, because
    the astronomical classification of SNe~Ia rests on the absence of
    hydrogen-features in the spectra of these events. The hydrogen
    stripped off from the companion will have rather low
    velocities. It may thus be detectable in nebular spectra, if
    abundant enough. For the single-degenerate scenario it is
    therefore of critical importance that the mass of stripped
    material is sufficiently low to be still consistent with the
    observations.

    There have been a few attempts to search for hydrogen in nebular spectra
    of SN Ia. \citet{mattila2005a} studied nebular spectra of SN
    2001el. From modeling them, they derived an upper limit of $0.03\,
    \mathrm{M}_{\odot}$ of solar
    abundance material at velocities lower than $1000\, \mathrm{km}\,
    \mathrm{s}^{-1}$. Recently, \citet{leonard2007a} studied
    nebular spectra of SN~2005am and SN~2005cf. Based on the same model
    as \citet{mattila2005a}, he estimated $\lesssim$$0.01\,
    \mathrm{M}_{\odot}$ of hydrogen material for both objects.
    Recently also hydrogen has been detected indirectly by \citet{patat2007a}
    in circumstellar material of SN 2006X.

    An alternative to this approach of observationally constraining
    the nature of the progenitor system is to directly search for
    the former companion star of the single-degenerate scenario in
    the remnants of historical galactic SNe~Ia. Such a search has been
    carried out in the remnant of Tycho Brahe's supernova of 1572 by
    \citet{ruiz-lapuente2004a}, who claimed the
    identification of the binary companion. The star in question is a
    slightly evolved solar-type star, that moves with a radial velocity of
    $-108 \, \mathrm{km\ s^{-1}}$ relative to the sun. It also has an atypical large tangential velocity
    of about $90 \, \mathrm{km\ s^{-1}}$. A significantly larger velocity of the
    star compared to neighbours is expected as a result of the disappearing binary
    orbit.
    Other stars observed in the same area
    with similar distances move only with average radial velocities of about
    $-20$ to $-40 \, \mathrm{km\ s^{-1}}$, with a velocity dispersion of about
    $20 \, \mathrm{km\ s^{-1}}$.

    On the theory side, \citet{marietta2000a} presented two-dimensional hydrodynamical
    simulations of the impact of SNe~Ia on their companions. They found
    that $0.15 \, \mathrm{M}_{\odot}$ were stripped from a Roche-lobe
    filling MS companion. This would rule out a MS-WD system
    for the SN~Ia analysed by \citet{leonard2007a}, if it was representative. Recently,
    \citet{meng2007a} pointed out that considering the effect of the
    mass transfer phase on the companion star may change the result
    significantly. They studied the impact of SNe~Ia on different
    companion stars analytically. In contrast to \citet{marietta2000a},
    who assumed the structure of single MS stars for the companion,
    \cite{meng2007a} evolved it through the binary evolution phase before
    the explosion of the WD. They found at least $0.035\,
    \mathrm{M}_{\odot}$ of stripped hydrogen for the
    companion. However, this result is only a lower limit, 
    since they did not include mass loss by vaporization from the hot
    surface of the star. Thus, taken at face value, the currently
    available theoretical studies constitute a strong case against
    MS+WD progenitor systems for SNe~Ia.

    The aim of the study presented here is to check and update the
    \citet{marietta2000a} calculations with the results of recent
    detailed binary evolution models. \citet{ivanova2004a} identified
    possible SN~Ia progenitors from a parameter study of  MS+WD
    binary evolution. There results are in agreement with other studies
    in this field \citep[e.g.][]{langer2000a, han2004a}. Based on these results,
    we present an exploration of the effect of the impact of a SN Ia on
    different MS companions by 3D hydrodynamical
    simulations. Section \ref{sec:code} summarizes the codes
    used. Section \ref{sec:tests} demonstrates that our approach reproduces the results of
    \cite{marietta2000a} and presents a resolution study. Section
    \ref{sec:results} discusses an exploration of different progenitor
    systems and
    Section \ref{sec:observ} derives observational implications of our results. A
    summary and an outlook conclude this work in Section
    \ref{sec:conclusion}. 

  \section{Modeling approach}
    \label{sec:code}
    Two different codes are employed in this work: one to construct
    the companion stars mimicking a binary evolution and the other to
    investigate the hydrodynamical impact
    of the supernova on the companion star.

    To evolve the companion stars we use the stellar evolution code
    GARSTEC of \cite{weiss2007a}. It evolves stars with a given mass and
    metallicity to a certain age and is used to construct a solar-type
    companion star similar to the ``HCV'' scenario of
    \citet{marietta2000a} (see Sect.~\ref{sec:tests}). The code also allows to include mass loss during
    the evolution but does not account for a binary evolution. In our
    study of a variety of progenitor models (see Sect.~\ref{sec:results}), we therefore rely on the parameters of the binary evolution study by
    \cite{ivanova2004a} to construct our companion stars.
    For each of the models we first set up a star that fits
    the parameters at the onset of the mass transfer phase. At this point, a
    constant mass loss rate is assumed and the stellar evolution is
    followed for the
    duration of the mass transfer period. The mass loss rate employed
    here corresponds to the
    mass loss 
    rate of the original binary models of \cite{ivanova2004a} averaged
    over the entire mass transfer phase.
    After following this phase, we obtain a stellar configuration that
    approximates the outcome of a realistic binary evolution. It is
    used to study the process of the interaction with the explosion
    ejecta once the binary WD undergoes a SN~Ia.

    The impact of the supernova ejecta is simulated using the smoothed
    particle hydrodynamics (SPH) code GADGET2 of \cite{springel2005a}.
    In order to set up the companion star here, we map the
    one-dimensional profiles of density, internal energy,
    and nuclear composition of the stellar evolution
    calculation to a particle distribution suitable for the SPH
    code. The mapping is done by transforming a uniform particle
    distribution to the given radial density profile. Details of
    this procedure will be explaind in \citep{pakmor2008b}.
    Using GADGET2, these companion stars were relaxed
    in a separate step for $1.0 \times 10^{4} \, \mathrm{s}$ to get
    rid of numerical artifacts that may have been introduced by the
    transformation (e.g.\ due to the random placement of the
    particles).

    A supernova was added to the simulation
    at a distance given by the orbital period of the binary system
    before the explosion. 
    The supernova was set up based on the W7
    model by \cite{nomoto1984a}. This one-dimensional model is
    well tested and provides a good fit to observations of standard SN Ia. It has a
    kinetic energy of $1.23 \times 10^{51} \, \mathrm{erg}$. At the time
    we add the supernova model to the simulation, it has reached already
    the phase of homologous expansion. The impact of the supernova
    ejecta and the following evolution of the ejecta and the companion
    star are simulated for about one hour. After this
    time, the companion star is already relaxing and its mass and velocity
    have reached constant values.

    The GADGET code was so far only used for cosmological simulations, with few
    exceptions \citep[e.g.,][]{morris2006a, morris2007a}.
    To be able to use it for a stellar astrophysics
    problem, a few extensions of the original implementation
    were necessary.
    The following basic setup was used in these simulations:
    \begin{itemize}
    \item
    The smoothing length is chosen such that a sphere of its radius
    encloses 
    80 neighboring particles. 
    \item
    The
    gravitational softening length
    equals the smoothing length. 
    \item
    All particles are given the same mass.
    \item
    Nuclear reactions are neglected in the simulations. This is
    justified since \cite{marietta2000a} showed 
    that the additional energy generated by shock-wave induced hydrogen
    burning is a marginal effect. The savings in computing time due to
    this approximation, however, are substantial.
    \end{itemize}
    Details of these changes are described in a
    separate publication \citep{pakmor2008b}.

    \begin{figure*}[!ht]
     \centering
     \includegraphics[height=0.8\textheight]{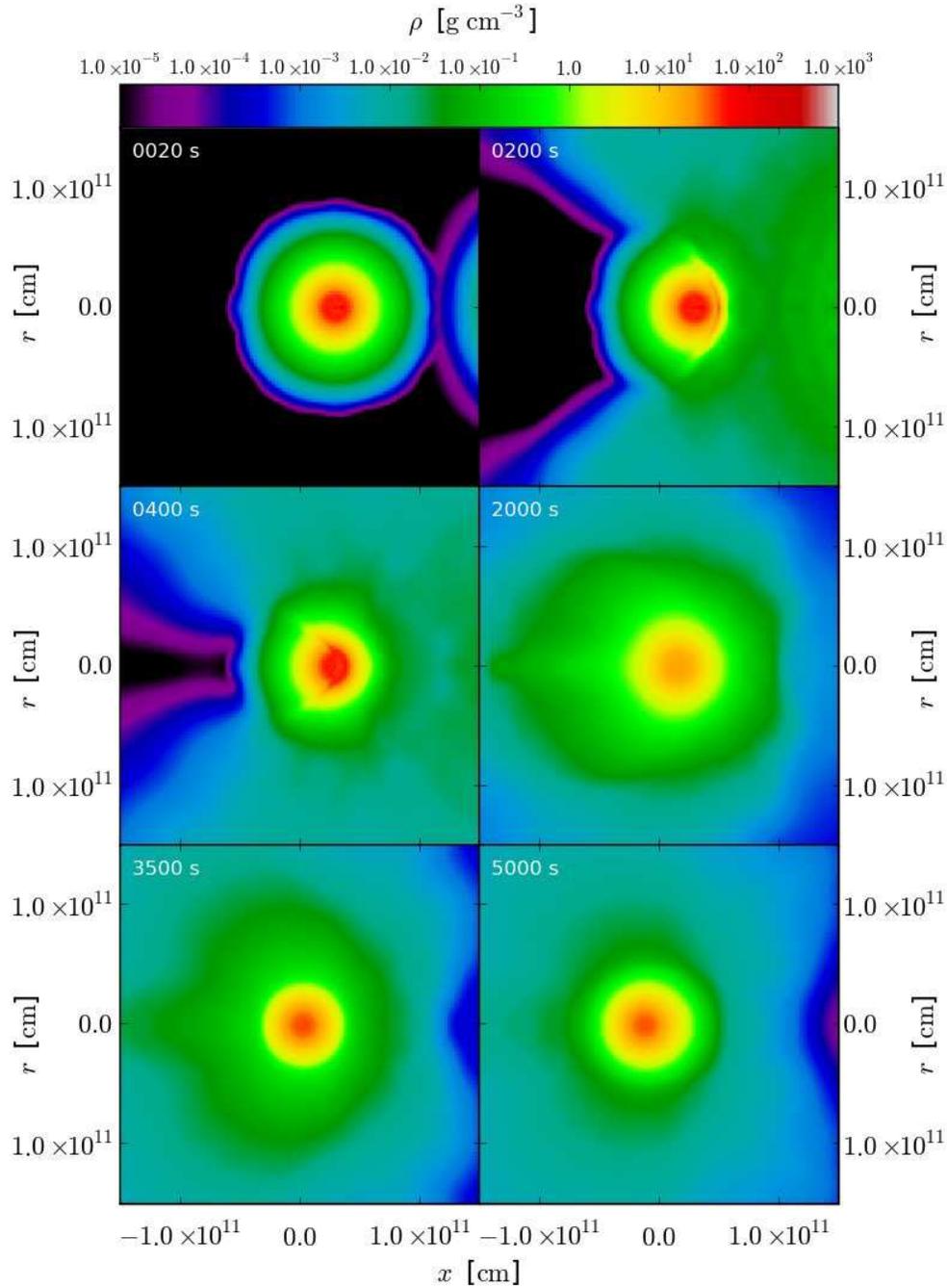}
     \caption{Snapshots of the evolution of the companion star in the
     HCV scenario. The plots use cylindrical
     coordinates. The radial coordinate is averaged over
     angle. Color-coded is the density.}
     \label{fig:marietta_evolution}
    \end{figure*}

  \section{Tests of implementation}
    \label{sec:tests}
    One of the obvious questions arising in our approach is whether
    the 3D SPH scheme applied here leads to the same results as the
    2D grid-based approach of \citet{marietta2000a}.
    This is tested by using the initial parameters of the HCV
    scenario of \citet{marietta2000a} in our setup. In the HCV scenario, the supernova
    is realized as a W7 model. The companion is a solar like
    $1.017\, \mathrm{M}_{\odot}$ main sequence star with a central
    hydrogen abundance of $0.58$. The separation
    between supernova and companion star at the time of the
    explosion is $2.04 \times 10^{11} \, \mathrm{cm}$. 

    Figure~\ref{fig:marietta_evolution} shows the typical evolution of the companion
    star in our simulations. Here, an example with a total of 235499
    SPH particles is illustrated starting out with the impact of the
    outermost supernova ejecta on the companion and ending when they
    have passed the star and it relaxes again. 
    The first image after $20\, \mathrm{s}$ shows the companion
    star at the instant when the first ejecta reach the companion from
    the right. In the second snapshot, taken after $200\, \mathrm{s}$,
    the ejecta have
    hit the companion star. A shock wave forms and starts to propagate
    through it.
    Another $200\, \mathrm{s}$ later (third snapshot of
    Fig.~\ref{fig:marietta_evolution}), the shock  wave reaches the
    center of the 
    star. In the fourth snapshot ($\sim$$2000\, \mathrm{s}$ after the
    explosion the 
    shock wave has crossed the star completely. Material is ejected on
    its far side. The last two images show the star
    shrinking and relaxing again.

    Qualitatively, this looks similar to the simulations of
    \cite{marietta2000a}. However, our results do not show the
    hydrodynamical instabilities they observed at the interface of
    supernova ejecta and the stripped material in the wake of the star
    (note that some mixing of companion star material into the
    supernova ejecta is realized, see Fig.~\ref{fig:abundance}).
    This difference is not surprising as SPH codes are known to
    suppress instabilities
     due to their large numerical viscosity \citep[see,
    however,][showing that grid-based codes may under some
    circumstances not reproduce the mixing better than SPH codes]{fryer2007a}. However, it is not a
    priori clear that this morphological difference significantly affects the effects and
    quantities we are interested in. The fundamental quantity upon 
    which we base our comparison is the bound mass of the companion
    star, or, equivalently, the mass stripped away from it by the
    supernova ejecta.

    As a first step, we perform a resolution study in order
    ensure a comparison with the \cite{marietta2000a} result
    based on a converged simulation.
    To test for
    numerical convergence we use different resolutions with $1.1 \times
    10^{5}$ to $4.7 \times 10^{6}$ SPH particles (counted for the
    entire setup, i.e.\ supernova and companion star, as the total
    mass of the configuration is equally distributed on the
    particles).

    \begin{figure}[h!]
     \centering
     \includegraphics[width=\linewidth]{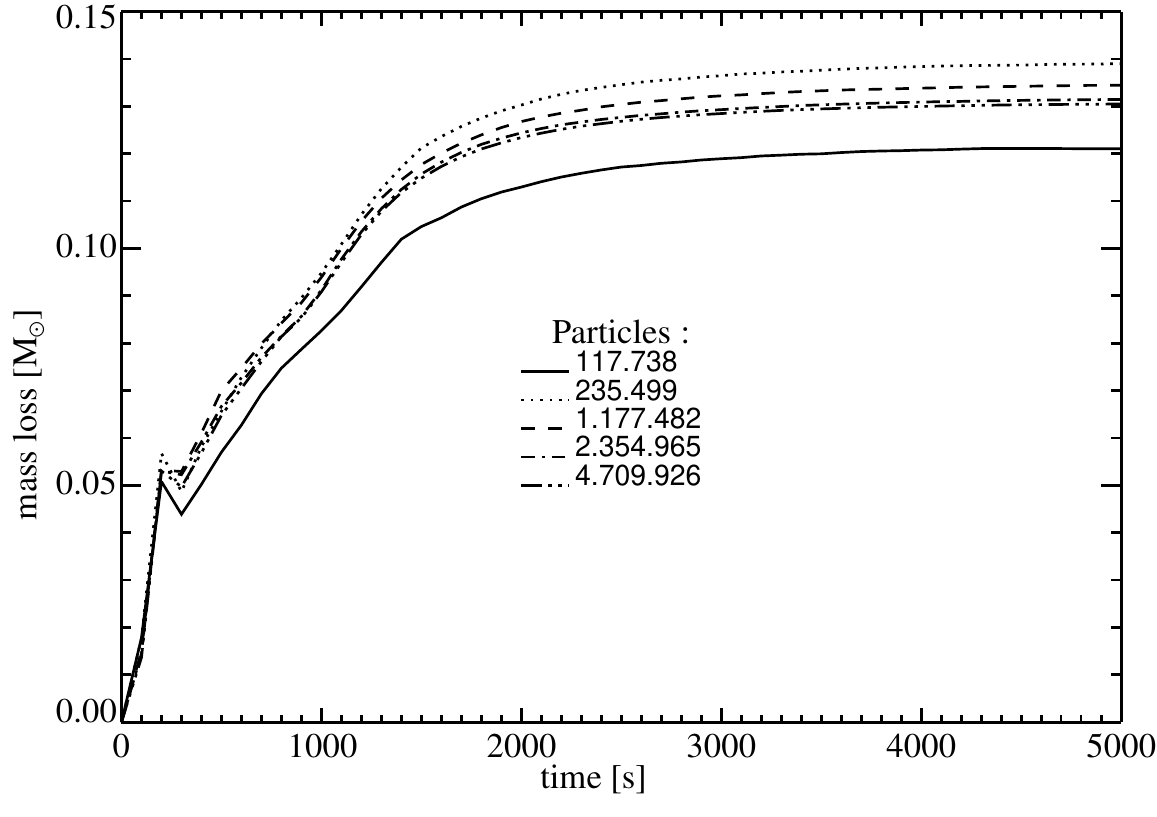}
     \caption{Mass loss of the companion star depending on the time
     after the supernova explosion for different
     resolutions.}
     \label{fig:marietta_resolution}
    \end{figure}

    Figure~\ref{fig:marietta_resolution} shows the evolution of the
    mass loss from the companion star with time for different
    resolutions. This value is calculated by subtracting
    the sum over the masses of all particles that are gravitationally
    bound to the star from its initial mass.  Whether or not
    a particle is bound to the star is decided by comparing its
    potential energy relative to the center of the star with the
    kinetic energy of its motion relative to the motion of the
    star. For this, the center 
    and velocity of the star are taken from the previous snapshot.
    Then the current center and velocity of the star are
    calculated from all bound particles. In principle, the new
    position of the star's center should 
    be used to recalculate which particles are bound to the
    star and this cycle should be iterated until convergence is reached.
    However, already the position and velocity of the star taken from
    the previous snapshot usually provide a sufficiently accurate
    approximation, and we thus forego the iteration.

    After $\sim$$3000\, \mathrm{s}$ the mass loss has stopped and the star has reached
    its final mass.
    There is a numerical artefact in the detection of the bound mass of the companion
    star around $\sim$$300\, \mathrm{s}$ after the explosion. For one snapshot the
    mass of the companion star seems to increase. This is due to a false detection of
    particles as unbound in the previous snapshot due to high radial velocities
    of particles, that point to the center of the star.

    \begin{table}[h]
      \begin{center}
      \begin{tabular}{rccccc}
        \hline \hline
        $n_\mathrm{star}$ & $n_\mathrm{tot}$ & $m_\mathrm{particle}$ & $m_\mathrm{stripped}$ & $m_\mathrm{remnant}$ & $v_\mathrm{kick}$ \\
        & & $[\mathrm{M}_\odot]$    & $[\mathrm{M}_\odot]$ &  $[\mathrm{M}_\odot]$ & $[\mathrm{km}\, \mathrm{s}^{-1}]$ \\
        \hline \\
          50000 &  117738 & $2.03 \times 10^{-5}$ & 0.126 & 0.891 & 101.1 \\
         100000 &  235499 & $1.02 \times 10^{-5}$ & 0.144 & 0.873 &  95.4 \\
         500000 & 1177482 & $2.03 \times 10^{-6}$ & 0.138 & 0.879 &  85.3 \\
        1000000 & 2352965 & $1.02 \times 10^{-6}$ & 0.135 & 0.882 &  80.3 \\
        2000000 & 4709926 & $5.09 \times 10^{-7}$ & 0.134 & 0.884 &  81.6 \\
      \end{tabular}
      \end{center}
      \caption{HCV resolution test:
      $n_\mathrm{star}$, $n_\mathrm{tot}$, and $m_\mathrm{particle}$ denote the number of particles
      the star is composed of, 
      the total number of particles in the simulation,
      and the particle mass, respectively. The results of the
      simulations are characterized by
      mass $m_\mathrm{stripped}$ stripped away from
      the companion star, its final mass $m_\mathrm{remnant}$, and its
      velocity relative to the supernova $v_\mathrm{kick}$. All values are
      taken $2 \times 10^4 \, \mathrm{s}$ after the explosion.\label{table:resolution}}
    \end{table}

    Table \ref{table:resolution} shows again some simulation
    properties for different resolutions after $2 \times 10^4
    \, \mathrm{s}$. Considering Table~\ref{table:resolution} and
    Fig.~\ref{fig:marietta_resolution}, we conclude, that
    this simulation is numerically converged when using more than
    $10^{6}$ particles. For the two simulations
    with 2.3 and 4.7 million particles, the graphs in Fig.~\ref{fig:marietta_resolution} are nearly identical. Note that the
    kick velocity converges slowlier than the stripped mass.

    Numerical convergence, however, does not necessarily imply
    consistency with the physical solution. To check our results in
    this respect, we compare them to previous
    results of \citet{marietta2000a}.
        With the highest resolution of nearly 5 million particles the
    stripped mass after $2.0 \times 10^{4} \, \mathrm{s}$ is $0.134\,
    \mathrm{M}_{\odot}$. This is quite close to the result of $0.15\,
    \mathrm{M}_{\odot}$ reported by \citet{marietta2000a}. 

    As carried out here, however, the comparison is not yet based on
    exactly the same assumptions. \citet{marietta2000a} performed their
    simulations on a finite computational grid and therefore mass was
    lost over the domain boundaries. This mass was always assumed to be
    unbound from the companion star. We therefore recalculate the
    stripped mass in our simulations by assuming for comparison all
    particles to be unbound that are outside a cylindrical box equal
    to the simulation volume of \citet{marietta2000a}. In this approach we find a stripped
    mass of $0.143\,
    \mathrm{M}_{\odot}$ for our highest resolved simulation. This
    result is in excellent agreement with the
    $0.15 \, \mathrm{M}_\odot$ of stripped mass found by
    \citet{marietta2000a}. 
    Moreover, the remnant star velocities
    at this time \citep[$85.7\, \mathrm{km}\, \mathrm{s}^{-1}$
    reported by][vs.\ our $81.4\, \mathrm{km}\,
    \mathrm{s}^{-1}$]{marietta2000a} agree 
    very well.

    We therefore conclude that our SPH approach is capable of
    capturing the main dynamical effects of the supernova impact on
    the companion star. The global quantities of the
    \citet{marietta2000a} study are reproduced down to the percent
    level. The differences in the occurrence of hydrodynamic
    instabilities are obviously a minor effect with respect to the
    overall results such as the stripped mass and the velocity
    of the companion star caused by the kick by the supernova
    ejecta. The instabilities suppressed in our SPH approach may,
    however, enhance the mixing between supernova
    ejecta and stripped material from the companion in reality.

  \section{Parameter studies}
  \label{sec:results}

    \begin{table*}[t]
      \begin{center}
        \begin{tabular}{lccccccc}
          \hline \hline
          & $M_\mathrm{c,i}$   & $M_\mathrm{c,f}$   & $\Delta t_\mathrm{tr}$ & $P_\mathrm{f}$ & $a_\mathrm{f}$        & $M_\mathrm{stripped}$ & $v_\mathrm{kick}$ \\
          Model & $[\mathrm{M}_\odot]$ & $[\mathrm{M}_\odot]$ & [yr]
          & [d]   & $[10^{11}\, \mathrm{cm}]$ & $(M_\odot)$    & $[\mathrm{km}\, \mathrm{s}^{-1}]$ \\
          \hline\\
          rp3\_28a & 2.8 & 0.6  & $7.7 \times 10^5$ & 1.7  & 5.21 & 0.032  & 52.8 \\
          rp3\_20a & 2.0 & 1.17 & $3.9 \times 10^6$ & 0.55 & 2.68 & 0.032  & 46.6 \\
          rp3\_20b & 2.0 & 1.25 & $2.0 \times 10^6$ & 1.08 & 4.26 & 0.0095 & 24.1 \\
          rp3\_25a & 2.5 & 1.37 & $1.7 \times 10^6$ & 0.51 & 2.62 & 0.058  & 60.5 \\
          rp3\_24a & 2.4 & 1.4  & $8.4 \times 10^5$ & 1.1  & 4.39 & 0.010  & 26.6 \\
          rp3\_20c & 2.0 & 1.46 & $2.6 \times 10^6$ & 1.44 & 5.29 & 0.012  & 17.0 \\
          \hline \\
        \end{tabular}
      \end{center}
      \caption{Parameters of the progenitor models and of the
          companion star after interaction with the supernova ejecta.\newline \emph{Values
          taken from \citep{ivanova2004a}:} masses of
      the companion star at the beginning of the mass transfer
      $M_\mathrm{d,i}$ and at the time of the explosion $M_\mathrm{d,f}$, length of
      the mass transfer period $\Delta
      t_\mathrm{tr}$,
      orbital period  $P_\mathrm{f}$, distance between white dwarf
      and its companion star just before the explosion $a_\mathrm{f}$;\newline
      \emph{Results of the simulations:} 
      mass stripped from the companion $M_\mathrm{stripped}$ and its
      kick velocity $v_\mathrm{kick}$, $5000\, \mathrm{s}$
      after the explosion. \label{table:binaries}} 
    \end{table*}

    Three major physical parameters of the progenitor system are
    expected to influence the dynamics of the supernova impact on the
    companion star: the kinetic energy of the supernova ejecta
    (powered by the thermonuclear burning in the explosion), the
    separation between supernova and companion, and structure of
    the companion star at the time of
    the explosion.

    In reality, of course, separation and companion structure are not
    independent, 
    but result from the characteristics of the original binary system
    and its evolution through 
    the mass transfer phase. Moreover, the dependence of the supernova
    explosion energy on other parameters is unknown and thus
    treated as an
    independent physical parameter in this study.
    In any system with Roch-lobe overflow, the solid angle under
    which the
    companion 
    star is seen from the white dwarf depends only on the
    mass ratio of its components.
    However, in order to determine
    how changing the distance affects the results it is treated here
    as an independent parameter. This is motivated partially by the
    possibility that the WD and the companion star disconnect shortly
    before the explosion, as suggested by \citet{patat2007a} for SN~2006X.

    Our main emphasis, however, will be on the structure of the
    companion star. In order to model the progenitor systems, we take
    parameters from the study of 
    \cite{ivanova2004a} and a typical Population I metallicity
    of $0.02$ for all stars. \cite{ivanova2004a} analyzed
    the evolution of possible SN Ia progenitor systems consisting of a
    WD and an evolved MS star. Detailed
    calculations of 65 different systems were carried out, varying the
    initial mass of
    both objects and the initial distance to cover the parameter
    space. From the resulting sample, they pick six representative
    models which are believed to be likely to evolve into a SN~Ia. These
    are the systems we select for our study. Their properties
    are listed in table \ref{table:binaries}.

    The three parameters are individually discussed in the following
    paragraphs. While the simulations testing the effects of the
    explosion energy and the binary separation were carried
    out with $2 \times 10^{5}$ particles in the companion star, the study based on the
    realistic progenitor system structure was set up with
    $2 \times 10^{6}$ and $4 \times
    10^{6}$ particles representing the companion. Therewith,
    the mass
    of the single particles and, as all particles had the same mass by
    construction, the
    number of particles representing the supernova explosion was
    fixed. 

    \subsection{Explosion energy}

      The influence of the
      supernova explosion energy on the interaction with the
      companion is studied on the basis of model rp3\_20a. All
      parameters but the supernova energy were
      kept constant with the values of the original model (see
      Table~\ref{table:binaries}).

      The kinetic
      energy of the supernova was varied in the range $0.8 \ldots 1.6
      \, \mathrm{B}\, (= 0.8 \ldots 1.6
      \times 10^{51}\, \mathrm{erg}$). The lower
      limit corresponds to lowest kinetic energies of simulated deflagration
      SNe~Ia. The upper limit is
      the maximum energy a SN~Ia can have, assuming that a
      Chandrasekhar-mass WD consisting of an equal-by-mass mixture of
      C and O burns completely to
      $\mathrm{Ni}^{56}$. The kinetic energy of the supernova ejecta
      $\mathrm{E}'_\mathrm{kin,SN}$ was adjusted by scaling
      the velocities $\vec{v}'$ of the supernova particles (originally
      representing the W7 model with
      $\mathrm{E}_\mathrm{kin,SN}^\mathrm{W7}$ and
      $\vec{v}^\mathrm{W7}$) according to 
      \begin{equation}
        \vec{v}' = \sqrt{
        \frac{\mathrm{E}_\mathrm{kin,SN}'}{\mathrm{E}_\mathrm{kin,SN}^\mathrm{W7}}
        } \cdot \vec{v}^\mathrm{W7}.
      \end{equation}
      This scaling preserves the homologous expansion ($v \propto r$) of the
      ejecta.

      \begin{figure}[!h]
        \centering
        \includegraphics[width=\linewidth]{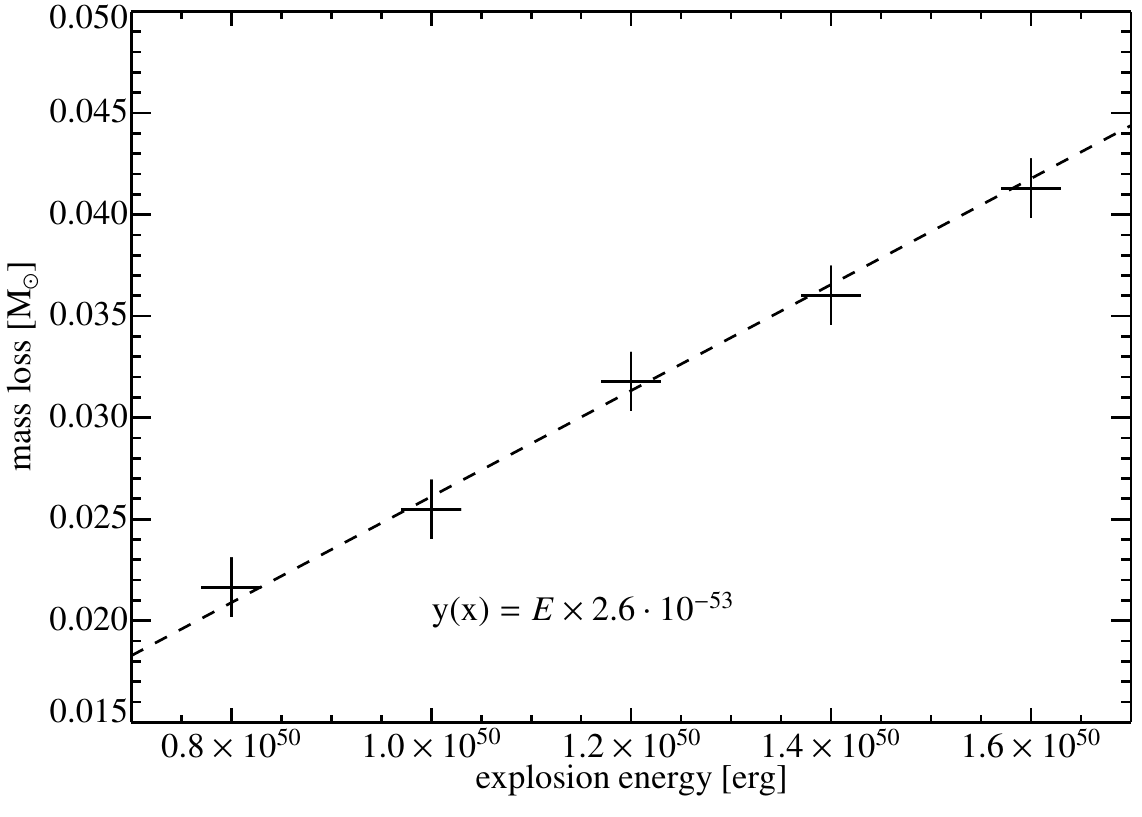}
        \caption{Stripped mass for different supernova energies in
        model rp3\_20a.} 
        \label{fig:param_snenergy}
      \end{figure}

      The stripped mass as a function of
      the supernova energy is shown in Fig.~\ref{fig:param_snenergy}. The relation is 
      linear in good approximation and can be fitted by
      \begin{equation}
        M_{\mathrm{stripped}} = 2.6 \times 10^{-2} \frac{E_{\mathrm{kin,SN}}}{10^{51}\mathrm{erg}} \, \mathrm{M}_\odot,
      \end{equation}
      assuming that the offset is zero (without this constraint the offset is only $1.3 \times 10^{-3}$).
      We emphasize that, although the functional form of the relation
      may be generic, the particular parameters of the fit apply to
      model rp3\_20a only. For different companion structures, for
      instance, the values of the parameters are expected to change. This should
      be kept in mind for the fits presented below as well.

      Note that the studied energy range covers
      only a factor of $2$ and therefore the stripped mass also changes
      only by a factor of $2$. The energy of the W7 model of $1.23 \times
      10^{51} \, \mathrm{erg}$ corresponds to an intermediate
      case. These results indicate that changes in the supernova
      energy have only a small effect on the stripped mass as compared
      to the other parameters discussed below.
      The kick velocity can be fitted by a power law as
      \begin{equation}
        \label{eq:energy_kick}
        v_{\mathrm{kick}} = 2.5 \times 10^{-22}
        \left(
        \frac{E_{\mathrm{kin,SN}}}{\mathrm{erg}} \right)^{0.55}
        \, \mathrm{cm}\,\mathrm{s}^{-1}.
      \end{equation}
      A simple argument for this behavior may be that velocity of the
      companion is mainly given by the momentum exchange from the supernova ejecta that
      hit the star. As the velocity of the ejecta increases with the
      square root of the kinetic energy, the velocity of the star
      should increase accordingly. The exponent in
      eq.~(\ref{eq:energy_kick}) is only slightly larger than $1/2$
      and thus this simple picture captures the process rather well. 

    \subsection{Distance}
\begin{figure}[!h]
        \centering
        \includegraphics[width=\linewidth]{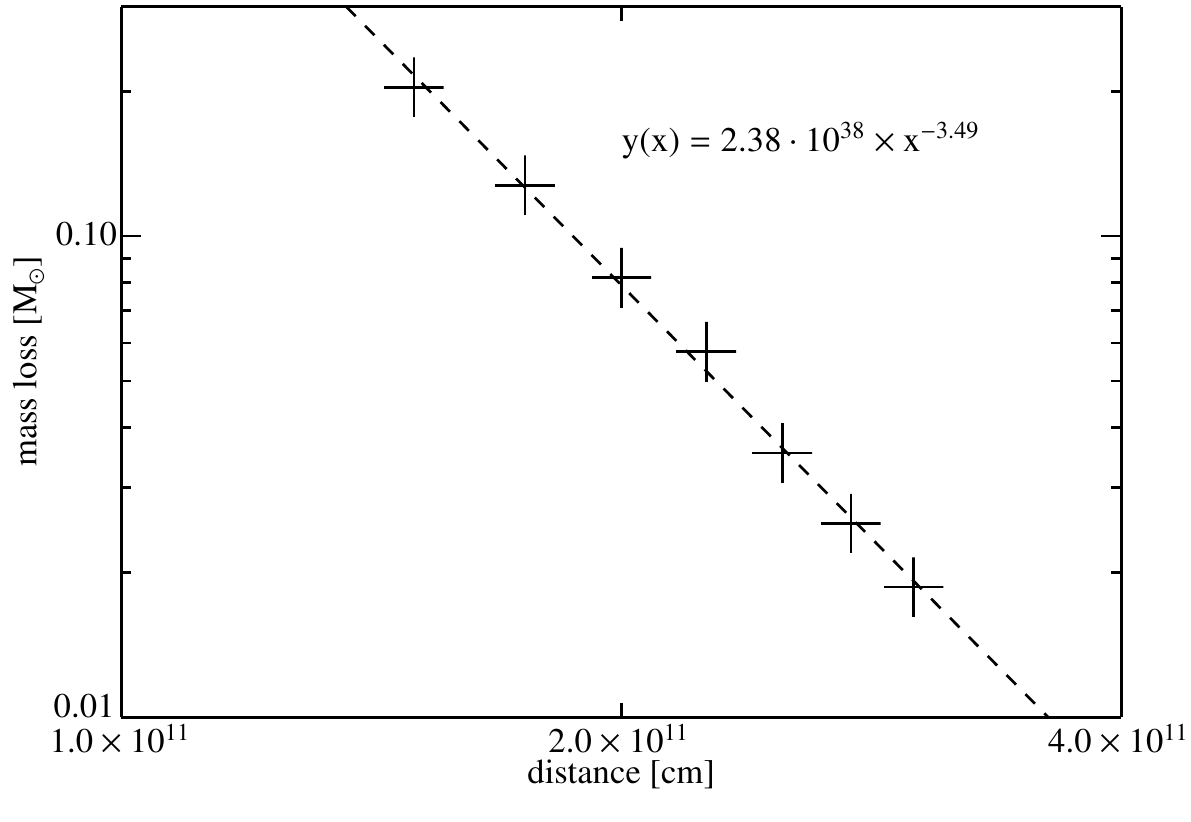}
        \caption{Stripped mass versus binary separation for
          model rp3\_20a.}
        \label{fig:param_distance}
      \end{figure}

      In order to test the influence of the separation, we again choose
      model rp3\_20a (see Table~\ref{table:binaries} for the
      parameters). Here, the distance $d$
      between WD and companion was varied in the range of $1.5 \ldots
      3 \times 10^{11} \, \mathrm{cm}$. All other
      parameters were kept constant. The relation between distance and
      stripped mass for this model is shown in Fig.~ \ref{fig:param_distance}. From the largest to the smallest
      distance the stripped mass increases by a factor of 10. The
      relation follows a power law in good approximation and can be
      fitted to
      \begin{equation}\label{eq:distance1}
        M_{\mathrm{stripped}} = 2.38 \times 10^{38} \left(
        \frac{\mathit{d}}{\mathrm{cm}} \right)^{-3.49}  \,
        \mathrm{M}_\odot.
      \end{equation}
      The kick velocity depending on the separation $d$ also
      follows a power law. It is approximately given by
      \begin{equation}\label{eq:distance2}
        v_{\mathrm{kick}} = 2.5 \times 10^{23} \left(
        \frac{\mathit{d}}{\mathrm{cm}} \right)^{-1.45}\,
        \mathrm{cm} \, \mathrm{s}^{-1}.
      \end{equation}

      The fraction of the supernova ejecta that hits the star scales with the
      inverse square of the distance treated as an
      independent parameter as discussed above. Remembering that the stripped mass seems
      to scale linearly with the explosion energy, one may expect a
      geometrical scaling as $M_{\mathrm{stripped}} \propto
      d^{-2}$. This does clearly not fit our results. It
      indicates that the connection is more complex, and the linear
      correlation between explosion energy and stripped mass is possibly
      only a result of the small energy range tested. In addition the simple scaling
      neglects that the companion star is not flat disc but a sphere. The whole
      process also seems to depend on the time evolution of the density of the ejecta
      hitting the star.

    \subsection{Companion star}
      \label{sec:param_comp}
      Table \ref{table:binaries} shows the stripped masses and the kick
      velocities of the remnants for the six different progenitor
      systems suggested by \citet{ivanova2004a}. The supernova model
      used in all models is the original W7 model with a 
      total kinetic energy of $1.23 \times 10^{51} \, \mathrm{erg}$.
      The
      variations in the stripped mass between the $2 \times 10^{6}$
      particles and the $4 \times 10^{6}$ particles simulations were less
      than a few percent indicating numerically converged results.

      The
      stripped masses range from $0.01 \, \mathrm{M}_\odot$ to $0.06\,
      \mathrm{M}_\odot$ for the different setups. This is significantly
      less than previous results reported by other authors
      \citep{marietta2000a, meng2007a}. Compared to the $0.15\,
      \mathrm{M}_\odot$ result
      of \cite{marietta2000a} we find a factor
      of $3$--$15$ less stripped material. 

      This deviation is attributed to the binary evolution of the
      progenitor we take
      into account in the present study. The
      binary evolution significantly affects the properties of both
      the companion star and the geometry of the binary configuration. 

\begin{figure}[!ht]
        \centering
        \includegraphics[width=\linewidth]{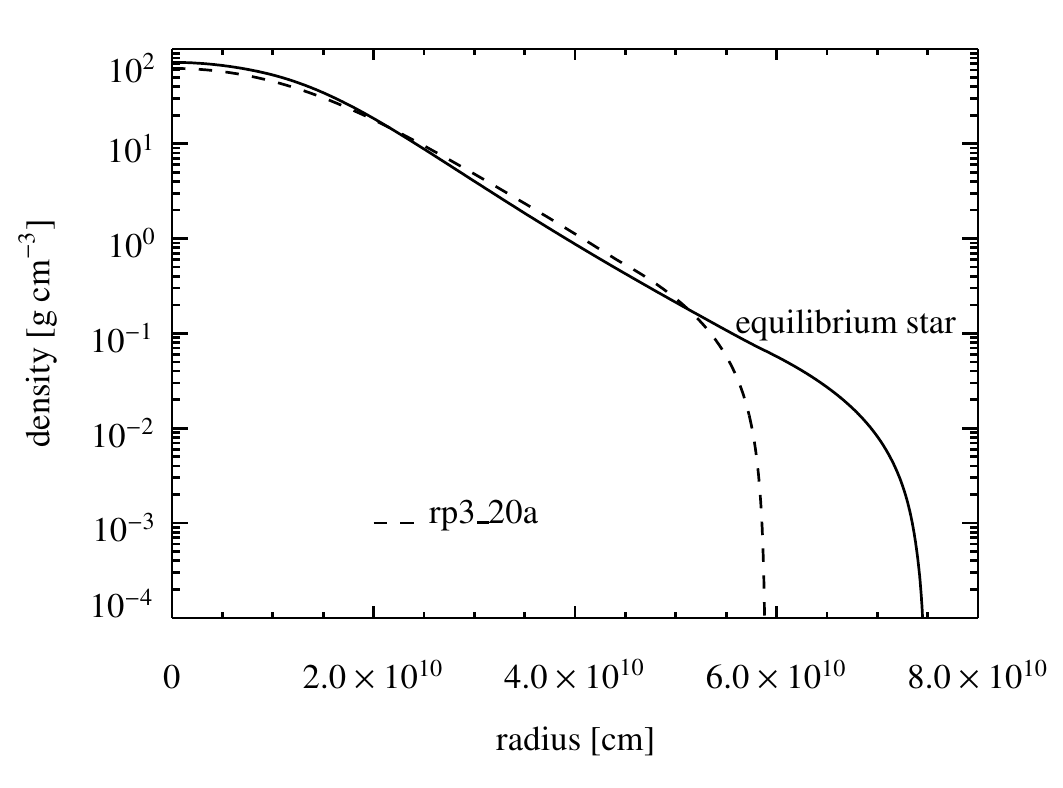}
        \caption{Comparison of density profiles of the companion star
        in model rp3\_20a at the time of the explosion and a single
        star with the same mass and nuclear age.} 
        \label{fig:binary_effects}
      \end{figure}

      The main
      effect on the companion star is illustrated in
      Fig.~\ref{fig:binary_effects}, where the density profile of the 
      companion star in model rp3\_20a at the time of the explosion is
      plotted in comparison with the density profile of a star in thermal
      equilibrium that was evolved as a
      single object to the same mass and nuclear age. The equilibrium
      star features 
      a much larger radius than the binary companion. This is a result
      of the mass 
      transfer phase in the binary system, that removes the outer
      layers and leads 
      to a more compact star. The mass loss occurs rather fast: the
      duration 
      of the mass transfer phase is about a factor of 10 less than the
      Kelvin-Helmholtz time-scale of the stars in our models.
      Therefore, the star is not able to adjust its structure to the
      loss of the outer 
      layers and it shrinks into a more compact object. Finding less
      stripped mass when
      taking into account the modification of the companion structure
      due to the 
      binary evolution is thus not surprising.

      In order to corroborate this interpretation, we set up  model
      rp3\_20a with a companion structure that corresponds to that of
      a star of equal mass in thermal equilibrium. In this model,
      $0.066\, \mathrm{M}_\odot$ are stripped from the
      companion by the supernova. This is a factor of 2 more than for
      the original rp3\_20a
      model. The difference can be explained by the larger, less bound
      envelope of the equilibrium star that can be stripped away more
      easily. Moreover, the larger radius causes an extended
      interaction area that also leads to a slightly larger kick
      velocity of $51.2\, 
      \mathrm{km}\, \mathrm{s}^{-1}$ for the equilibrium star
      companion model evolved in isolation as compared
      to the $46.6\, \mathrm{km}\, \mathrm{s}^{-1}$ of the binary star
      companion model. 

      Additionally, in the model we use the companion star is slightly more massive
      ($1.17\, \mathrm{M}_\odot$ at the time of the explosion) than
      the solar mass companion star of \cite{marietta2000a}. This leads to a slightly
      larger radius of the equilibrium star companion and therewith a larger separation
      distance and a decreased stripped mass.

      These effects together explain the significant
      difference between our results and the previous work of
      \cite{marietta2000a}, who used a companion with the structure of
      an isolated equilibrium star. We note, however, that our interpretation is
      not in agreement with the work of \cite{meng2007a}, who
      found a lower limit on the stripped mass of $0.035\, \mathrm{M}_\odot$ 
      taking into account binary evolution. This is probably
      be due to oversimplifying assumptions they made in their
      analytical treatment of the ejecta-companion star interaction.

      The kick velocities of the companion star after the impact of
      the supernova ejecta we find in our models vary from $17\, \mathrm{km}\,
      \mathrm{s}^{-1}$ to $61\, \mathrm{km}\, \mathrm{s}^{-1}$ (see Table~\ref{table:binaries}). 
      The velocities roughly increase with the size of the
      companion stars (as the cross section increases) and decrease
      with larger separation distances (as the transferred momentum decreases)
      and masses (for the same transferred momentum) of the companion stars.

  \section{Observational implications}

    Observations that may help to constrain the nature of the SN~Ia
    progenitor systems are the detection of hydrogen in the spectra of
    these events, signatures in the spectrapolarimetry data due to an
    asymmetric morphology of the ejecta as a result of the interaction
    with the companion star, and a direct observation of companions
    in supernova remnants. We discuss these possibilities in the light
    of the presented study.

    \label{sec:observ}
    \subsection{Hydrogen detection in SN~Ia spectra}
      Our results are consistent with the constraints on hydrogen in
      the ejecta as given by
      \cite{leonard2007a} and \cite{mattila2005a}. However, none of our
      models has stripped hydrogen mass far below the upper limit of
      \cite{leonard2007a} of $0.01\, \mathrm{M}_\odot$. As these
      limits result from a 
      non-detection of hydrogen in nebular spectra, our simulations
      predict hydrogen detection not far below these values.
      Conclusions on the validity of the single-degenerate scenario
      depend on whether the systems studied here are representative
      for 
      Chandrasekhar mass WD+MS SN~Ia progenitors and whether
      the observed events fall into this
      class. If both were true, lowering the
      observational upper limits of hydrogen in the ejecta of SNe~Ia
      by another order of magnitude would exclude this progenitor scenario.

      Yet it is important to note that a quite simple model was employed
      by \cite{leonard2007a} and \cite{mattila2005a} to constrain
      limits on the hydrogen mass from the observations. A more
      rigorous approach would be to use the results of hydrodynamical
      simulations such as presented here as an input for full
      radiative transport 
      calculations. From these calculations it will be possible
      to predict whether hydrogen lines should be visible in the spectra
      and how strong they should be at a given epoch. This issue will be
      addressed in a forthcoming study.

    \begin{figure*}[!ht]
     \centering
     \includegraphics[height=0.8\textheight]{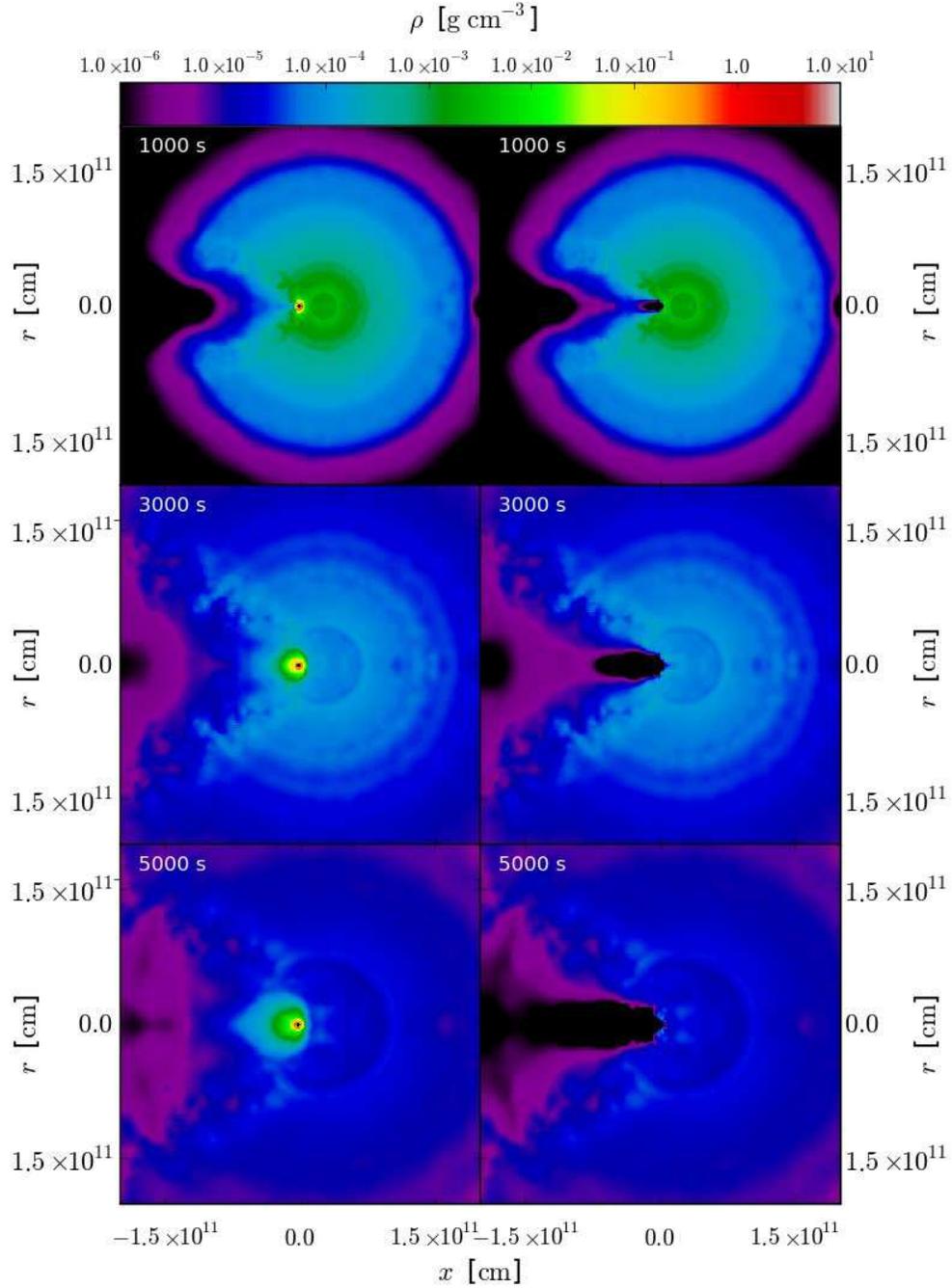}
     \caption{Snapshots of the evolution of the supernova ejecta in
     the rp3\_20a scenario. The left column shows supernova ejecta and
     the companion star. In the right column all material of the companion is cut out, leaving
     only the supernova material. The plots use cylindrical
     coordinates. The radial coordinate is averaged over
     angle. Color-coded is the density.}
     \label{fig:ejecta}
    \end{figure*}

    \subsection{Hole in the ejecta}
      The supernova ejecta do not only affect the companion star, but
      are also affected themselves by the impact. Fig.~\ref{fig:ejecta}
      shows the material after the 
      impact. The left column shows both the supernova ejecta and the
      companion star material including the hydrogen stripped from the
      star. The right column
      shows only the material that was part of the supernova ejecta at
      the beginning of the simulation. 

      The supernova ejecta that were
      spherically symmetric in the beginning are clearly asymmetric
      after the impact. In the wake of the companion star, a
      cone-like hole in the supernova ejecta is visible. To some degree it is
      filled with material that is ejected from the companion star. At
      the borders of this hole, the supernova ejecta are slightly
      denser, because the material missing in the hole was
      transfered there. The opening angle of the cone-like hole is about $45^\circ$
      (see the top row of Fig.~\ref{fig:ejecta} showing the ejecta
      $1000\,\mathrm{s}$ after the explosion). This result
      is consistent with the findings of \cite{marietta2000a}. However,
      the area in which the supernova ejecta are affected by the
      impact on the companion star is as large as $90^\circ$ (see
      middle and the lower row of Fig.~\ref{fig:ejecta}
      showing the density of supernova ejecta after $3000\, \mathrm{s}$ and
      $5000\, \mathrm{s}$).

\begin{figure*}[!ht]
     \centering
     \includegraphics[width=0.8\textwidth]{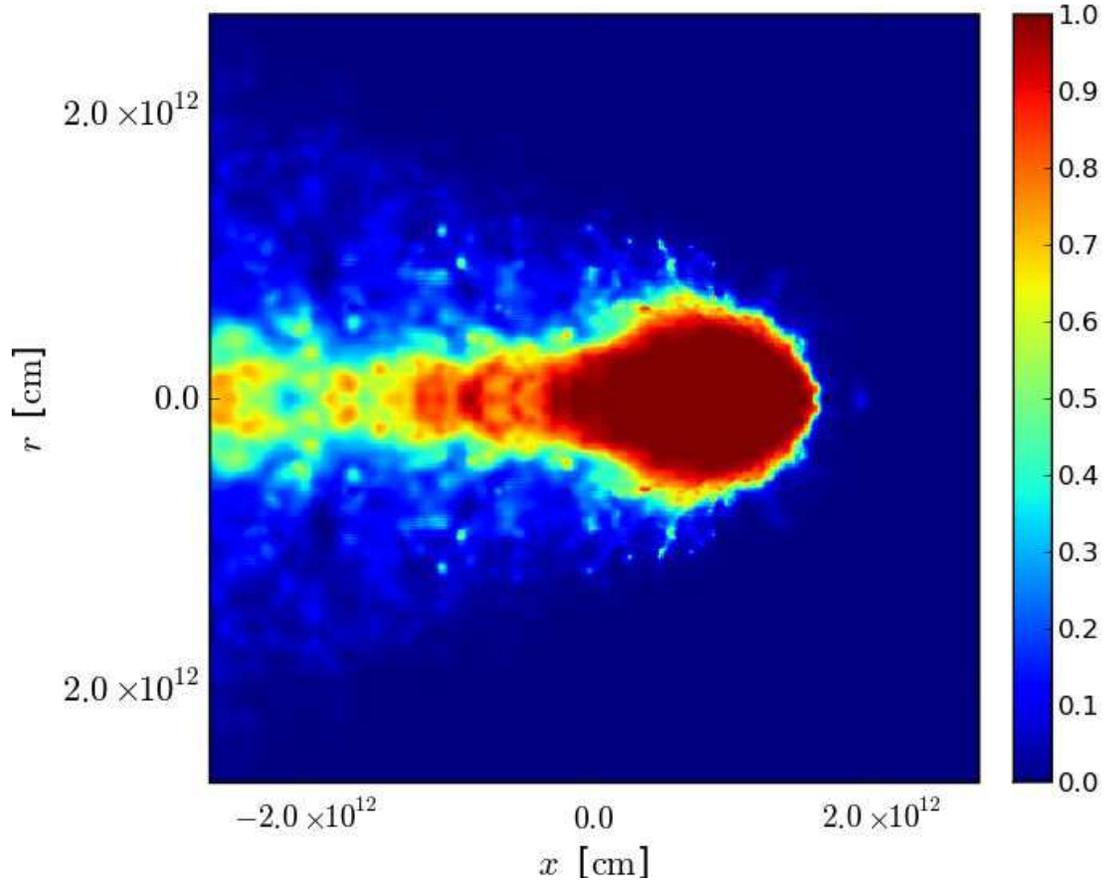}
     \caption{Snapshot of simulation rp3\_20a taken $5000\,
     \mathrm{s}$ after the explosion. 
     Color-coded is the of the relative amount of material
     originally belonging to the companion star (red) with respect to
     the total material (blue corresponds to supernova material).
     Cylindrical coordinates with the radial coordinate averaged over
     angle are used.}
     \label{fig:abundance}
    \end{figure*}

      A more detailed view of how the supernova ejecta are mixed with the material stripped
      from the companion star is given in figure
      \ref{fig:abundance}. It shows for the 
      the same setup
      the relative amount of material that originally belonged
      to the companion star with respect to the total amount of material.

      Qualitatively, the structure of the ejecta resembles the simple
      model used by \cite{kasen2004a} to explore the effect of a hole
      in the ejecta on spectra and luminosity of SN Ia. With it he was
      able to reproduce observed spectrapolarimetry observations of SN Ia.

    \subsection{Identifying companion stars in supernova remnants}
      The remaining companion star should have a different velocity than
      its surrounding stars. This velocity is determined by the orbital
      velocity of the star at the moment the white dwarf explodes
      and of the kick it gets from the impact of the supernova
      ejecta. The former is perpendicular to the connecting line between
      the white dwarf and the companion star. For our models it ranges
      between
      $130 \, \mathrm{km}\, \mathrm{s}^{-1}$ (rp3\_20c) and $380\, \mathrm{km}\, \mathrm{s}^{-1}$ (rp3\_28a).
      The latter velocity is a result of the impact of the ejecta and
      is therefore
      aligned with the connecting line. In our simulations it reaches values from $17 \mathrm{km\ s^{-1}}$
      to $50\, \mathrm{km}\, \mathrm{s}^{-1}$. Therefore, the orbital velocity clearly
      dominates the velocity of the companion star relative to the
      center of the supernova remnants. The velocity of the star Tycho
      G which was 
      identified as progenitor of Tycho Brahe's supernova by
      \citet{ruiz-lapuente2004a} features a modulus of its spatial velocity of
      $136\, \mathrm{km}\, \mathrm{s}^{-1}$. It thus 
      falls into lowest part of the range we find in our simulations. The
      properties of Tycho G are thus consistent with the predictions of our
      models.

  \section{Conclusions}
    \label{sec:conclusion}
    We studied the impact of the ejecta of SNe~Ia on
    main sequence companion stars in the context of the single-degenerate
    Chandrasekhar-mass scenario with hydrodynamical
    simulations. To this end, the (cosmological)
    GADGET2 smoothed particle hydrodynamics code was adapted to the stellar problem and employed
    in numerical simulations. It was shown that this SPH-based approach
    is capable of reproducing previous results obtained with a
    grid-based 2D scheme by
    \cite{marietta2000a}. A resolution
    study indicated that with a few million particles the
    simulations yield numerically converged results.

    We showed that the mass stripped from
    the companion star by the impact of the ejecta depends on the
    their kinetic energy and the binary separation. While the
    latter affects the mass of the stripped material significantly,
    the former has only a minor effect on it. This
    is due to the fact that the supernova explosion energy can only
    vary in a relatively narrow range given the restricted amount
    of fuel available for the nuclear energy generation. 

    The SPH approach was used to analyze the impact in a number of more
    realistic progenitor models than those employed in previous
    studies. For these, the companion stars were constructed with the stellar evolution code
    GARSTEC mimicking binary mass transfers with the parameters given
    by \citet{ivanova2004a}. 
    In the hydrodynamical impact simulations, we found about one order of magnitude
    less hydrogen material stripped off the companion by the impact of
    the supernova than predicted by previous studies. The
    main reason for this difference is a modified, more compact stellar structure
    of the companion star in combination with a resulting variation in the separation
    distance of the progenitor system. In particular the more compact state of the companion
    impedes the mass loss in the impact.

    The reduced amount of hydrogen mixed into the ejecta of the
    supernova as predicted by our simulations leads to an
    agreement with observational studies of SN~Ia nebular
    spectra \citep{mattila2005a,leonard2007a}. This removes the former
    disagreement between the available 
    observations and simulations of the WD+MS progenitor system. Thus,
    to current knowledge, such a progenitor scenario is admissible in
    the context discussed here.

    However, since the hydrogen masses predicted by our simulations
    are not far below the current observational upper limits,
    it may be possible in the near future to confirm or reject the studied
    progenitor scenario by either detecting hydrogen in SNe~Ia or
    lowering the limits by another order of magnitude. A stringent way
    of analysis would be to calculate synthetic
    spectra directly from the presented simulations and to compare the
    results with observations. This will be tackled in a forthcoming
    study. 

    \begin{acknowledgements}
      We thank Hans Ritter for helpful discussions on the binary
      evolution of the progenitor systems. Volker~Springel and
      Klaus~Dolag gave invaluable support with the numerical
      implementation into \citet{ivanova2004a}the framework of the 
      the GADGET2 code. The research of F.K.R.\ is supported
      through the Emmy Noether Program of the German Research Foundation (DFG;
      RO~3676/1-1). Additional support was came from the Cluster of
      Excellence EXC~153 ``Origin and Structure of the Universe'' and
      from the Transregional Collaborative Research Center TR~33.
    \end{acknowledgements}

  \bibliographystyle{aa}

\end{document}